\documentstyle[preprint,aps]{revtex}

\begin{document}

\draft

\title{Radial equation for an accretion flow  driven by   Poynting flux}

\author{Hyun Kyu Lee\footnote{e-mail : hklee@hepth.hanyang.ac.kr}}

\address{Department of Physics, Hanyang University, Seoul 133-791, Korea \\
and \\ Asia Pacific Center for Theoretical Physics, Pohang 790-784, Korea}

\maketitle

\begin{abstract}
Using a toy model of a two dimensional accretion disk, we discuss further the radial equation
of the accretion flow dominated by the Poynting flux.    Assuming  the  force-free condition is
valid around the accretion disk, the relation between the fluid angular velocity  of the disk,
$\Omega_D$,  and the  angular velocity of the magnetic field lines,
 $\Omega_F$, is also discussed.

\pacs{PACS numbers: 97.10.Gz, 97.60.Lf}

\end{abstract}

\narrowtext

\section{Introduction}

The model of Poynting flux  has been suggested for describing the
astrophysical jets\cite{blandford}\cite{lovelace}\cite{BZ} and
discussed also for gamma ray bursts
\cite{lwb}\cite{lbw}\cite{li}\cite{vp} and the evolution of a
black hole-accretion disk system \cite{lk}\cite{lp}.

Recently the effect of  Poynting flux on accretion flows has been
formulated relativistically in the background of Kerr geometry for
the rotating balck hole at the center  \cite{hkl}.  A simplified
model of a two-dimensional accretion disk located on the
equatorial plane is assumed as in the non-relativistic formulation
\cite{blandford}. To see the effects of the Poynting flux
transparently it is also assumed to be non-viscous and
non-radiative, although the realistic accretion disk is much more
complicated\cite{acgl}.

In this brief note, we reformulate the two dimensional accretion
model suggested in ref. \cite{hkl}, where the time- and toroidal
part of the energy-momentum conservation has been discussed to
calculate the accretion rate and the Poynting flux. The radial
part of the energy-momentum conservation is  discussed in detail
to see how the angular velocity of magnetic field, $\Omega_F$, is
related to that of the accretion flow, $\Omega_D$.

The structure of the electromagnetic field with an accretion disk
which lies on $\theta= \pi/2$ plane can be reproduced by assigning
the surface charge and current on the plane in addition to the
bulk charge and current distributions outside the disk.

We can define a conserved current as\cite{damour}
\begin{eqnarray}
{\cal J}^{\mu} = J_+^{\mu} \Theta(\pi/2 -\theta) +J_-^{\mu}
\Theta(\theta -\pi/2) + j^{\mu}, \end{eqnarray} where
$J_{\pm}^{\mu}$ is the bulk current
 densities( subscripts  $+$ and $ ~ -~$ stand
 for the upper and lower half space respectively)
which appear in  the Maxwell equation,
\begin{eqnarray}
F^{\mu\nu}_{\pm\,\,; \nu} = 4\pi J_{\pm}^{\mu},\label{maxwellb}
\end{eqnarray} which terminates on the disk. and $j^{\mu}$ is the
surface current density defined on the two dimensional accretion
disk.  Using the current conservation, ${\cal J}^{\mu}_{\,\,;\mu}
=0$
 we can identify  the surface current on the disk as
 \footnote{$[A] \equiv A_+ - A_-$. The electric and magnetic
field are defined for ZAMO(Zero Angular Momentum
Observer)\cite{TPM}.}
\begin{eqnarray}
j^{\mu} =\frac{1}{4\pi} [F^{\theta \mu }]\delta(\theta- \pi/2).
\end{eqnarray}

Following  the standard procedure the surface  charge density,
$\sigma_e$, and surface current density, $\vec{K} $, on the disk
can be obtained  by
\begin{eqnarray}
\sigma_e &=&  -\frac{ [E^{\hat{\theta}}]}{4\pi}, \label{eE}
\\  \vec{K} &=& \frac{1}{ 4\pi} [\vec{B}]\times \hat{\theta},\label{kbad}
\end{eqnarray}
  which are  the Gauss' law and the Ampere's law on
the surface of the accretion disk.   The discontinuity of the
fields at the disk plane due to the surface charge and current
leads to the Dirac-delta singularity given by

\begin{eqnarray}
\partial_{\theta} E^{\hat{\theta}}   &=&  4 \pi \sigma_e ~ \delta(\theta -\pi/2),\\
\partial_{\theta} \vec{B} ~  &=& 4 \pi \vec{K} \times \hat{\theta} ~ \delta(\theta
-\pi/2).
\end{eqnarray}
 on the accretion disk.

\section{Conservation of energy - momentum tensor}

The conservation of the energy momentum tensor given by
\begin{eqnarray} T^{\mu \nu}_{~~~;\mu} = 0 \label{tmn0} \end{eqnarray}
has four equations, one for each $\nu$ .   For the steady and
axial symmetric case, there are two Killing vectors and therefore
correspondingly the energy and the angular momentum(azimuthal
component) of the system are conserved.  The relevant combinations
of equations of $\nu = 0, ~ \phi$ are responsible for the energy
and the angular momentum conservation. The radial component of the
equations, $\nu = r$, determines the orbital motion of the fluid.

 As in ref.\cite{hkl} we simplify the accretion
disk to be non-viscous, cool and non-radiative with $p=T=0$ for
the purpose of investigating the effect of the magnetic field
transparently. Then the stress-energy tensor of the matter is
given simply by
\begin{eqnarray}
T^{\mu\nu}_{m} = \rho_m u^{\mu}u^{\nu}. \end{eqnarray} It is also
assumed that there is no mass flow in the direction perpendicular
to the disk: $u^{\theta} =0$.

\subsection{Energy and angular momentum conservation}

The Killing vectors, $\xi^{\mu}$ in  $t-direction$ and $\eta$ in
$\phi-direction$
\begin{eqnarray}
\xi^{\mu} &=& (1,\, 0,\, 0,\,0), ~~ \eta^{\mu} = (0,\, 0,\, 0,\,1)
 \end{eqnarray}
define the energy flux ${\cal E}^{\mu}$  and angular momentum flux
${\cal L}^{\mu}_{(\phi)}$ respectively
\begin{eqnarray}
{\cal E}^{\mu} &=& - T^{\mu\nu}\xi_{\nu}, ~~ {\cal
L}^{\mu}_{(\phi)} = - T^{\mu\nu}\eta_{\nu},
\end{eqnarray}
which  satisfy
\begin{eqnarray}
{\cal E}^{\mu}_{~~ ;\mu} = 0, ~~ {\cal
L}^{\mu}_{(\phi);\mu}=0.\label{eac}
 \end{eqnarray}

It can be  decomposed into two parts:  contributions from
matter($m$)and electromagnetic ($EM$) parts:
\begin{eqnarray}
{\cal E}^{\mu} &=& {\cal E}_{m}^{\mu} + {\cal E}_{EM}^{\mu}, ~
{\cal L}^{\mu}_{(\phi)} = {\cal L}^{\mu}_{m(\phi)} + {\cal
L}^{\mu}_{EM(\phi)} \end{eqnarray}

Since we are interested in a steady and axisymmetric case, the
derivatives with respect to $t$ and $\phi$ vanishes and hence $r-$
and $\theta-$ components  of the flux are only considered:
\begin{eqnarray}
{\cal E}^{\mu}_{~~ ;\mu}
&=&\frac{1}{\sqrt{-g}} \{\partial_{r}(\sqrt{-g}{\cal E}^{r}) +
\partial_{\theta}(\sqrt{-g}{\cal E}^{\theta})\},\\
{\cal L}^{\mu}_{{(\phi)} ;\mu} &=&\frac{1}{\sqrt{-g}}
\{\partial_{r}(\sqrt{-g}{\cal L}^{r}_{(\phi)}) +
\partial_{\theta}(\sqrt{-g}{\cal L}^{\theta}_{(\phi)})\}.
\end{eqnarray} where
\begin{eqnarray}
{\cal E}_{m}^{r} &=& -\rho_m u_0 u^{r}, ~~~~   {\cal
E}_{m}^{\theta} =0\\ {\cal E}_{EM}^{r}&=&
-\frac{1}{4\pi}\frac{\sqrt{\Delta}}{\rho}(\alpha E^{\hat{\theta}}
+
\beta \varpi B^{\hat{r}})B^{\hat{\phi}}, ~~
 {\cal E}_{EM}^{\theta} = -\frac{1}{4\pi \rho}(-\alpha E^{\hat{r}} +
\beta \varpi B^{\hat{\theta}})B^{\hat{\phi}},\\
 {\cal L}^{r}_{m(\phi)} &=& \rho_m
u_{\phi}u^{r}, ~~  {\cal L}^{\theta}_{m(\phi)}=0 \\{\cal
L}^{r}_{EM(\phi)} &=& -\frac{1}{4\pi}\frac{\varpi
 \sqrt{\Delta}}{\rho} B^{\hat{\phi}}B^{\hat{r}},
 ~~{\cal L}^{\theta}_{EM(\phi)}= -\frac{1}{4\pi}
 \frac{\varpi}{\rho} B^{\hat{\phi}}B^{\theta}. \end{eqnarray}

Since we assume an idealized thin disk, we do not expect any
substantial  radial-flow of electromagnetic energy and angular
momentum compared to the flow of the accreting matter which has a
delta-function type singularity, for example,
\begin{eqnarray}
{\cal E}_{m}^{r} = -\rho_m u_0u^{r} = -\frac{\sigma_m}{r}u_0u^{r}
\delta(\theta-\pi/2) \end{eqnarray}  However  the radial flow
$\partial_{r}(\sqrt{-g}{\cal E}^{r}_{EM})$ has no delta-type
singularity such that a radial flow can be ignored in this thin
disk approximation.

On the other hand  the flux in the $\theta $ direction has a
singularity on the disk due to the discontinuity of the
electromagnetic fields.  Keeping  singular parts only , we obtain
\begin{eqnarray}
\partial_r(u_0) \dot{M}_+  +   2\pi r K^r(-\alpha E^{\hat{r}} +
\beta \varpi B^{\hat{\theta}}) =0,
\label{ec1}\\
\partial_r(u_{\phi}) \dot{M}_+  + 2 \pi r K^r \varpi B^{\hat{\theta}} =0
\label{amc1}, \end{eqnarray} for energy and angular momentum
conservation respectively. The mass accretion rate derived from
the rest-mass conservation is given by\cite{acgl}
\begin{eqnarray}
\dot{M}_+ &=& - 2 \pi r \sigma_m u^r \label{mplus}\end{eqnarray}
Eq.(\ref{ec1}) and (\ref{amc1}) have been  derived from the
condition that there should be no accumulation of energy and
angular momentum in the stationary accretion flow driven by
Poynting flux as discussed in ref. \cite{hkl}.

\subsection{Radial equation}

The radial component of the energy-momentum conservation
\begin{eqnarray}
T^{\mu r}_{~~~,\mu} = 0 \label{tr}\end{eqnarray} is useful to
investigate the angular velocity of the fluid. The matter part can
be written as
\begin{eqnarray}
T^{\mu r}_{m ~~ ;\mu}
 &=& -\rho_m g^{rr} u^0
\partial_r(u_{\phi})(\frac{u^{\phi}}{u^0} + \frac{\partial_r
u_0}{\partial_r u_{\phi}})
 \end{eqnarray}
 And the electromagnetic
part of the radial equation is given by
\begin{eqnarray}
T^{\mu r}_{EM~ ;\mu} = \frac{1}{\sqrt{-g}} \partial_{\mu}
(\sqrt{-g} T_{EM}^{\mu r}) + \Gamma^r_{\mu\lambda}
T_{EM}^{\mu\lambda}. \end{eqnarray} Since only singular parts  are
relevant for the thin accretion disk, we get
\begin{eqnarray}
T^{\mu r}_{EM~ ;\mu} &\rightarrow& \frac{1}{\sqrt{-g}}
\partial_{\theta} (\sqrt{-g} T_{EM}^{\theta r})\\
&\rightarrow& - \frac{\sqrt{\Delta}}{ r^2} (\sigma_e E^{\hat{r}}
-K^{\hat{\phi}} B^{\hat{\theta}}) \delta(\theta -\pi/2),
\end{eqnarray} where $\rightarrow$ indicates the procedure of keeping
singular parts only.

After a short straight forward calculations,   the radial equation, eq.(\ref{tr}), can be
written by
\begin{eqnarray}
-\frac{\sigma_m}{r} g^{rr}u^0\partial_r (u_{\phi})(\Omega_D + \frac{\partial_r u_0}{\partial_r
u_{\phi}}) -\frac{\sqrt{\Delta}} {r^2} (\sigma_e E^{\hat{r}} -K^{\hat{\phi}} B^{\hat{\theta}})
 =0
 \label{rad1}\end{eqnarray}
where $\Omega_D\equiv  u^{\phi}/u^0$ is the angular velocity of
the fluid in an accretion disk. The last term corresponds to the
Lorentz force exerted by the surface charge and current.

In the absence of the external magnetic field,   the radial equation can be satisfied with
\begin{eqnarray}
\Omega_D = -  \frac{\partial_r u_0}{\partial_r u_{\phi}},
\end{eqnarray} which is one of the characteristics of Keplerian
orbit around the black hole.  We can see that the orbits of the
particle in the disk under external electromagnetic field deviate
from the Keplerian orbit as expected.  Hence, the trial ansatz,
$\Omega_D = \Omega_K$,  used in the literature as a first
approximation for the order-of-magnitude calculations, might not
be valid for a disk dominated mainly by the Poynting flux.

\section{Force-free Magnetosphere}

In this section  we assume that there is a sufficient ambient plasma around the disk to
maintain the force-free magnetosphere\cite{blandford}\cite{love}. The force-free condition for
the magnetosphere with the current density $J^{\mu}$ is given by
\begin{eqnarray}
F_{\mu\nu}J^{\mu} = 0. \label{ff}
 \end{eqnarray}

With  force-free condition, magnetic field lines are spiraling
rigidly to form magnetic surfaces\cite{TPM}  with the angular
velocity $\Omega_F$ and we get
\begin{eqnarray}
\vec{E} = -v_F^{\hat{\phi}} \times \vec{B}^P, ~~ v_F^{\hat{\phi}}=
\frac{\varpi}{\alpha}(\Omega_F +
\beta) \hat{\phi}\label{epa}
\end{eqnarray} where $v^{\hat{\phi}}_F$ is the toroidal velocity observed by ZAMO.
Then the toroidal surface current density is given by\footnote{It is consistent with the
non-relativistic treatment, Eq.(3.34) multiplied by $\alpha$ and Eq.(3.36) in
\cite{blandford}.}
\begin{eqnarray}
K^{\hat{\phi}} = \sigma_e / v_F \end{eqnarray}
 which seems to look strange when compared to the
naive expectation
\begin{eqnarray}
K^{\hat{\phi}} \sim \sigma_e  v_F. \end{eqnarray}  It is simply because the surface charge
defined by the boundary condition is not the charge relevant for the surface current and $v_F$
is not the velocity of the fluid  in the disk.

Now  the Poynting flux is given by
\begin{eqnarray}
 {\cal E}_{EM}^{\theta}  =
\Omega_F {\cal L}^{\theta}_{(\phi)} \label{erhatd2},
\end{eqnarray} and eq.(\ref{ec1})  becomes
\begin{eqnarray}
~~~~~~ \partial_r(u_0) \dot{M}_+  -   2\pi r K^r\varpi \Omega_F
 B^{\hat{\theta}} =0 \label{emc1}. \end{eqnarray}

 When compared with eq.(\ref{amc1}), the angular velocity of the
magnetic surface in a force-free magnetosphere is found to be
related to the fluid velocity in the following way
\begin{eqnarray}
\Omega_F = - \frac{d u_{0}/dr}{d u_{\phi}/dr} \label{dudr}.
\end{eqnarray}  It is inetersting to note that it  has
no explicit dependence on the electromagnetic field.

One of the interesting cases is $u_0$ and $u_{\phi}$ are those of
the Keplerian orbit\cite{hkl} where
\begin{eqnarray}
-\frac{d u_{0}/dr}{d u_{\phi}/dr} = \Omega_K \end{eqnarray} and
therefore
\begin{eqnarray}
\Omega_F = \Omega_K \label{ofk}\end{eqnarray} However  it is subject to the radial equation,
Eq.(\ref{rad1}), in which the naive suggestion, Eq.(\ref{ofk}), is not guaranteed in a simple
way as discussed in the previous section.  In fact the radial equation provides   more
information rather on the relation between $\Omega_D$ and $\Omega_F$.

 Eq.(\ref{rad1}) can be written
under the force-free condition as
\begin{eqnarray}
-\frac{\sigma_m}{r} g^{rr}u^0\partial_r u_{\phi}(\Omega_D -\Omega_F) -\frac{\sqrt{\Delta}}
{r^2} (\sigma_e v^{\hat{\phi}}_F -K^{\hat{\phi}}) B^{\hat{\theta}}
 =0. \label{rad3}
 \end{eqnarray}
The second part represent the Lorentz force in the radial direction and it should be noted that
the force-free condition does not apply on particle motions in the disk.  Hence one can see
that  the angular velocity of the field lines is not the same as that of the disk. It is quite
contrary to the conventional assumption,  $\Omega_F = \Omega_D$, for the highly conducting
accretion disk.  It is rather similar to the case with black hole in a force-free
magnetosphere,
 where the angular velocity of the black hole, $\Omega_H$,  is different from $\Omega_F$\cite{TPM}.
 This equation shows how $\Omega_F$ in the force-free magnetosphere can be
determined by $\Omega_D$.

Naively  one can  consider the case where the toroidal current $K^{\hat{\phi}}$ is entirely due
to the surface charge density, $\sigma_e$,  rotating with the angular velocity of the fluid,
$\Omega_D$ . Then the toroidal current is given by
\begin{eqnarray}
K^{\hat{\phi}} = \sigma v_D^{\hat{\phi}}, v_D^{\hat{\phi}}
 = \frac{\varpi}{\alpha}(\Omega_D + \beta) \label{kzamo} \end{eqnarray}
 where $v_D^{\hat{\phi}}$ is the local velocity of the matter  observed by ZAMO.
Then Eq. (\ref{rad3}) becomes
\begin{eqnarray}
\{-\frac{\sigma_m}{r} g^{rr}u^0\partial_r u_{\phi}
+\frac{\sqrt{\Delta}} {r^2} \sigma \frac{\varpi}{\alpha}
B^{\hat{\theta}}\}(\Omega_D -\Omega_F)
 =0 \label{rad4}
 \end{eqnarray}
which seem to show  that the angular velocity of the magnetic field line is the same as  the
fluid angular velocity,
\begin{eqnarray}
\Omega_F = \Omega_D.\end{eqnarray} However it does not hold every where on the disk but only
for the particular section of the accretion disk where
\begin{eqnarray}
 v^{\hat{\phi}}_D(r) = v^{\hat{\phi}}_F(r) = c.
\end{eqnarray}
which might not be useful in the realistic situation.  Of course
it is due to an assumption that the surface current density is
entirely  due to the rotating surface  density defined by the
boundary condition.

\section{Summary}

In this work,  a model of the two-dimensional accretion flow
driven by Poynting flux discussed in \cite{hkl} is reformulated to
discuss the radial equation of the acrreting matter. Due to the
Lorentz force, the angular velocity of the fluid motion,
$\Omega_D$ deviates from the Keplerian angular velocity $
\Omega_K$.  With  force-free magnetosphere, we get a relation
between $\Omega_F$ and fluid motion: $ \Omega_F = - \frac{d
u_{0}/dr}{d u_{\phi}/dr} $. However it is found that there is a
difference between $\Omega_F$ and $\Omega_D$ as for the case of
rotating black hole in a force-free magnetosphere.  In practical
applications with accretion disk, the discussions can be very much
simplified by assuming $\Omega_D \sim \Omega_K $ and/or $\Omega_F
\sim \Omega_D$.  Hence it would be  very interesting to estimate
the deviations from those ideal equalities  and it remains as
future numerical works to do.

It is also worth noting that the recent numerical work\cite{love}
demonstrates the possibility of a Poynting flux dominated
accretion flow. Although a non-relativistic formulation is used,
it suggests a possibility that at least a part of the accretion
flow can be driven dominantly by the Poynting flux.

\vskip 1cm

The author would like to thank Hongsu Kim for useful discussions.
This work was supported by Korea Research Foundation Grant
(KRF-2000-015-DP0081).


\begin{thebibliography}{}



\bibitem{blandford}  R.D. Blandford, MNRAS {\bf 176}, 465(1976)


\bibitem{lovelace} R.V.E. Lovelace,  Nature {\bf 262}, 649(1976)

\bibitem{BZ} R.D. Blandford   and Z.L. Znajek, MNRAS  {\bf 179},
433(1977)


\bibitem{lwb} H.K. Lee,  R.A.M.J. Wijers and  G.E.  Brown, Phys.
Rep. {\bf 325}, 83(2000)

\bibitem{lbw} H.K. Lee,  G.E. Brown and   R.A.M.J. Wijers, ApJ
{\bf 536}, 416(2000)

\bibitem{li} L.-X. Li,  Phys. Rev. {\bf  D61}, 084016(2000)

\bibitem{vp} M.H.P.M. van Putten, Phys. Rep. {\bf 345}, 1(2001)

\bibitem{lk} H.K. Lee and H.K. Kim, J. Korean Phys. Soc. {\bf 36},
188(2000)

\bibitem{lp} L.-X. Li, B. Paczynski, ApJ {\bf 534}, L197(2000)

\bibitem{hkl} H.K. Lee, Phy. Rev. D64, 043006(2001)


\bibitem{acgl} see,  for example,  M.A. Abramowicz,   X.-M. Chen,
M. Granath and J.-P. Lasota,    ApJ {\bf 471}, 762(1996)

\bibitem{damour} T. Damour,  Phys. Rev. D {\bf 18}, 3598(1978)



\bibitem{TPM} K.S. Thorne, R.H.  Price,  and D.A. Macdonald, ``Black Holes;
The Membrane Paradigm" (Yale University Press, New Haven and
London, 1986)

\bibitem{love} G.V. Ustyugova, R.V.E. Lovelace, M.M. Romanova,
H. Li and S.A. Colgate,   ApJ  {\bf 541}, L21(2000)


\end{thebibliography}
\end{document}